\documentstyle[12pt]{article}     % Paper without sections

\def\monthyear{\ifcase\month\or
  January\or February\or March\or April\or May\or June\or
  July\or August\or September\or October\or November\or
December\fi
  \space\number\year}

\def\up#1{\leavevmode \raise.16ex\hbox{#1}}

\def\slash#1{{\mathpalette\c@ncel{#1}}}

\renewcommand{\baselinestretch}{1.17}
\newcommand{\gapproxeq}{\lower
.7ex\hbox{$\;\stackrel{\textstyle >}{\sim}\;$}}
\newcommand{\lapproxeq}{\lower
.7ex\hbox{$\;\stackrel{\textstyle <}{\sim}\;$}}

\newcounter{appendice}

\def\thefiglist#1{\section*{Figure Captions\markboth
 {FIGURE CAPTIONS}{FIGURE CAPTIONS}}\list
 {Figure \arabic{enumi}.}
 {\settowidth\labelwidth{Figure #1.}\leftmargin\labelwidth
 \advance\leftmargin\labelsep
 \usecounter{enumi}}
 \def\baselinestretch{1.1}\@normalsize
 \def\newblock{\hskip .11em plus .33em minus -.07em}
 \sloppy}

\newcommand{\be}{\begin{equation}}
\newcommand{\ee}{\end{equation}}
\newcommand{\bea}{\begin{eqnarray}}
\newcommand{\eea}{\end{eqnarray}}
\newcommand{\bean}{\begin{eqnarray*}}
\newcommand{\eean}{\end{eqnarray*}}

 \newcommand{\bsea}{\begin{subeqnarray}}
\newcommand{\esea}{\end{subeqnarray}}

\def\thefiglist#1{\section*{Figure
Captions\markboth
{FIGURE CAPTIONS} {FIGURE
CAPTIONS}}\list
{Figure \arabic{enumi}}
{\settowidth\labelwidth{Figure
#1.}\leftmargin\labelwidth
\advance\leftmargin\labelsep
\usecounter{enumi}.}
\def\newblock{\hskip .11em plus .33em
minus -.07em}
\sloppy}

\def\thefiglist#1{\section*{Figure Captions\markboth
{FIGURE CAPTIONS} {FIGURE CAPTIONS}}\list
{Figure \arabic{enumi}.}
{\settowidth\labelwidth{Figure #1.}\leftmargin\labelwidth
\advance\leftmargin\labelsep
\usecounter{enumi}}
\def\newblock{\hskip .11em plus .33em minus -.07em}
\sloppy}

\begin{document}
\begin{titlepage}
\begin{flushright} RAL-92-022\\
\end{flushright}
\vskip 2cm
%\null
%\vskip .5cm
\begin{center}
{\bf\large Effective Heavy Quark Theory }\\
\vskip 1cm
{\bf \large F E Close\footnote{FEC @ UKACRL}}\\
Rutherford Appleton Laboratory,\\
Chilton, Didcot, Oxon,
OX11 OQX, England\\
and\\
{\bf \large Zhenping Li\footnote{ZPLI @ UTKVX}}\\
Dept of Physics\\
University of Tennessee, Knoxville, TN 37996-1200, USA
\end{center}
\begin{abstract}
We show how heavy quark effective theory, including $\frac{1}{M_Q}$
corrections,
may be matched onto dynamical quark models by making a specific choice
of
$K_\mu, m$  and $v_\mu$ in the $p_\mu = mv_\mu + K_\mu$
expansion.
We note that Wigner rotations of heavy quark spins arise at $O(p^2/m^2$)
in
non-relativistic models but at $O(\Lambda_{QCD} /M_Q)$ or
$O(velocity-transfer)$ in HQET and so are
necessary for a consistent treatment.
\end{abstract}
\vskip3in
  May 1992
\end{titlepage}

There is currently considerable experimental and theoretical interest in
hadrons
that contain a single heavy quark.  On the experimental side, flavour
changing
weak decays such as $B\rightarrow D^* e\nu$ provide essential input to
completing the elements in the CKM matrix as part of a larger strategy of
unravelling the source of CP violation.  The problem for theory is to
extract
these CKM elements at {\bf quark} level from experiments that involve
{\bf
 hadrons}.
Historically this has relied on model calculations where a heavy quark is
bound
in a heavy hadron: these have the virtue of being applied to $b,c,s$ or
even
 $u,d$
flavours  but suffer from model dependent assumptions and, in the case of
$u,d$
flavours, considerable dispute as to reliability and self-consistency.
Recently
there has been interest in an approach (heavy quark effective theory,
HQET)
which exploits newly discovered symmetries of QCD that apply to Green
functions of heavy quarks that are nearly on-shell [1,2].  In HQET the states
depend on the velocity of the heavy quark and involve an expansion
around
infinite quark mass with $v$ held finite.

In practice $m_Q\neq \infty$.  This is perhaps a reasonable
approximation
for
$b$, but for $c$ and $s$ one needs to understand corrections of
$O(\Lambda_{QCD}/m_Q)$.  However, there is some arbitrariness in how
one
defines the variables.  The states in HQET depend upon the velocity $v$ of
the
heavy quark which is conserved in the absence of external interactions
(the
``velocity superselection rule" [2]).  Ref 3 has noted some ambiguities in
procedure concerning the relation between the field of velocity $v_\mu$
and
a
heavy-light meson that contains this heavy quark.  For example, quoting
from
ref 3 ``Is the meson's momentum $mv_\mu$ or does it differ by some
amount
$K_\mu$ of order of the hadronic scale $\Lambda$?  Is $m$ the mass of
the
meson or the quark?"   We propose a physically motivated choice that
enables
immediate contact with the formalism for describing composite systems
e.g.
heavy
quark in a heavy hadron.  This highlights that the {\bf light} quark(s)
 in a heavy-light system play a nontrivial dynamical role and are not
``spin-inert"\footnote{See e.g. the comments on p 348 and 356 of ref 16}
even though,
 at first sight,
only the heavy quark appears to be involved directly in flavor
changing decays such as $B\to D^* l\nu$.  Furthermore we shall
see that this spin activity survives even
in the leading order when $m_Q\to \infty$.

First we review the HQET expansion to $O(1/M_Q)$.

A heavy quark $Q$ is interacting with light degrees of freedom whose four
momenta are of order $\Lambda_{QCD}$ and much smaller than the
heavy
quark mass, $m_Q$.  In HQET one writes
\be
p^\mu_Q = m_Q v^\mu +K^\mu
\ee
where $K^\mu$ is a residual momentum, small compared to $m_Q$.
The
heavy quark spinor is written [2]
\be
Q=e^{-i m_Qv\cdot x} [h^{(Q)}_v + \chi^{(Q)}_v]
\ee
where $h$ survives as $m_Q v \to \infty$ and $\chi$ is an $O(K/m_Q)$
correction.

The decomposition is defined explicitly by [2]
\be
\not{v} h^{(Q)}_v = h^{(Q)}_v , \not{v} \chi^{(Q)}_v = -\chi^{(Q)}_v
\ee
and it is straightforward to verify that the equations of motion
then impose the constraints, to $O(1/m_Q^2)$
\be
\frac{\rlap /{\hskip -0.035 in K}}{2m_Q} h_v =\chi_v \quad ;\quad
v\cdot
K =0
\ee

This is a perfectly legitimate and actively researched
procedure when applied to
the Lagrangian for a single heavy quark.  However, many
practical applications deal specifically with a heavy
quark $(Q)$ within a many-quark system, in which case
there are kinematic correlations among the $Q$ and other constituents
that are manifested in processes involving recoil, e.g.
 $H_1(Q_i)\to H_2(Q_j) l \nu$ or in  Compton scattering.
 The requirements of gauge invariance, current
conservation and Lorentz invariance {\bf constrain}
the definition of dynamical variables such that, for example,
the low energy theorems of Compton scattering are
satisfied in an arbitrary frame for a many
quark system [4,6,8].

This is our point of departure.

We suggest that it may be useful to impose these
 constraints at the outset and thereby identify
a {\bf particular} choice of $v$, $m$ and $K$ from
 the more general set of possibilities consistent
 with Eq. 1. Our aim is to choose the variables so
 that the $\chi$, $h$
spinors of the HQET refer respectively to the motion
of the heavy quark relative to
the heavy hadron and the overall motion of the system with
$v_\mu$ identified
as the hadron's velocity; the $K_\mu$ will be a function of the relative
momentum of the heavy quark in the hadron.  Our prescription, which
is consistent with the general
precepts of HQET, leads to  immediate and consistent
insertion of the heavy quark in a heavy hadron, hence
 isolating an ``effective heavy quark theory".

At first sight this appears trivial. It is tempting to exploit
the familiar separation of overall,$\vec P$, and relative, $\vec K$,
momentum variables for a non-relativistic composite system
\bea
\left. \begin{array}{lll}
\vec{p}_Q &=&\frac{m_Q}{M} \vec{P} +\vec{K} \\
E_Q &=&\frac{m_Q}{M}E + \frac{\vec{K}\cdot\vec{P}}{M}
\end{array}\right\}
\eea
and to compare with eq.(1)  where $M$ is the hadron's mass
and $v_{\mu}=(E/M;\vec P/M)$ is its four-velocity.
However, this is wrong.
 The reason why, and the correct solution, can be seen if
 we first recall where Eq. 5 comes from.  If $p_\mu =(\omega, \vec{k}$) in
the
hadron $\vec{P} =0$ frame, (where $k=O(\Lambda_{QCD}$)) then
after a Lorentz boost to a frame with $\vec{P} \neq 0$, the variables are
\bea
\left. \begin{array}{lll}
\vec{p}_Q &=& \frac{\omega_Q\vec{P}}{M} +\vec k +
\frac{\vec k\cdot \vec{P}}{M(E+M)} \vec{P}\\
E_Q &=& \frac{E}{M} \omega_Q +\frac{\vec k\cdot \vec{P}}{M} \\
\sigma(\vec{v}) &=& \vec{\sigma} + \frac{(\vec k\times
\vec{P})\times
\vec{\sigma}}{2M(E+M)} \end{array}\right\}.
\eea
If $m, M\rightarrow \infty$ with $k/M,P/M\rightarrow 0$ then eq (5)
obtains;
furthermore in this limit $\sigma (\vec{v}) \equiv \vec{\sigma  }$ and
the
spin
is trivial.
However,
in HQET, $m_Q\rightarrow\infty$ with $v$ {\bf held finite}. As
 $v^\mu
=P^\mu/M$ is the hadron's velocity, then eq (6) becomes (to
$O(\Lambda^2/m^2_Q))$
\bea
\left. \begin{array}{lll}
\vec{p}_Q &=& m_Q\vec{v}+\vec k
+\frac{\vec{k}\cdot\vec{v}}{1+v_0}
\vec{v}\\
E_Q &=& m_Qv_0 +\vec k\cdot \vec{v} \\
\sigma(\vec{v}) &=& \vec{\sigma} + \frac{(\vec k\times
\vec{v})\times
\vec{\sigma}}{2m_Q(1+v_0)}
\end{array}\right\}
\eea
The equations for $(\vec{p}_Q, E_Q)$ have the form of eq (1); notice that
$\vec{\sigma}(\vec{v})$ is no longer trivial and that $v$ is understood to
be
the {\bf hadron's} velocity which,as is clear from eq.7, {\bf differs} from
the heavy quark velocity in general. As we shall see later, this is important
when matching with explicit composite wavefunctions.

If $m_Q$ in eq (1) is identified as the quark mass (technically, its energy
$\omega =\sqrt{m^2+\vec k^2}$ in the hadron's rest frame) then the
residual
momentum $K^\mu$ has the frame dependent form
(from eq 7)
\be
K^\mu \equiv k^\mu +\frac{\vec{v}\cdot\vec k}{1+v_0}  v^\mu
\ee
Notice that the constraint equation $K\cdot v=0$ implies that $k\cdot v=-
k_0$
 or,
equivalently, that
\be
\frac{\vec{v}\cdot \vec k}{1+v_0} =k_0.
\ee
Alternatively, if one wishes to identify $K^\mu$ as the residual
momentum $k^\mu$ of the heavy quark in the hadron's rest frame, then
it is
the mass parameter that must be regarded as  frame dependent in order to
realise
eqs(7)
\be
p^\mu_Q = m^{eff}_Q v^\mu +k^\mu \quad ;\quad m^{eff}_Q
=\omega_Q+
\frac{\vec{v}\cdot\vec k}{1+v_0}.
\ee

Now we shall study the decomposition of the heavy quark spinor, eqs(2-4),
in terms of these variables.

We choose to define $h_v$ to be independent of $m_Q$ and hence
\be
h_v= \sqrt{\frac{1+v_0}{2}}
\pmatrix{1\cr
\frac{\vec{\sigma}\cdot\vec{v}}{1+v_0}\cr}
\ee
With $K^{\mu}$ defined at eq(8) and
taking care to include, consistently, the $O(\vec{v}/m)$
 spin-rotation (eq 7c) and the constraint at eq(9),
 the $O(k/m_Q)$ correction to the heavy quark spinor is
\be
\chi_v\equiv \frac{\rlap /{\hskip -0.035 in K}}{2m_Q} h_v
=\sqrt{\frac{1+v_0}{2}}
\pmatrix{\frac{\vec{\sigma}\cdot\vec{v}}{1+v_0}\cr 1\cr}
\frac{\vec{\sigma}\cdot\vec k}{2m_Q}.
\ee
where the $\vec{\sigma}$ operator acts on spinor states as
defined in the hadron  $\vec{v}=0$ frame.

Note the ordering of $\vec{\sigma}\cdot\vec{v}
\vec{\sigma}\cdot\vec{k}$
that arises in the HQET expansion, eq(12). We shall now show that this
matches
the spinor representation of a composite system if the latter is constructed
in the system's rest frame and then boosted to arbitrary velocity.

  In the hadron rest frame we write
formally
\be
H(\vec{v}=0) =Q(\vec{v}=0;\vec k) \otimes S(\vec v=0; -\vec k)
\ee
where $Q(\vec k), S(-\vec k)$ refer to the heavy quark and spectator
system
respectively with equal and opposite three momenta (``relative
momenta").
To
$O(k^2/m^2)$ the spinor of the heavy quark is
\be
Q(\vec{v}=0;\vec k) =
\pmatrix{1\cr
\frac{\vec{\sigma}\cdot\vec k}{2m_Q}\cr}
\ee
(we are neglecting binding energy at present).Upon boosting to a frame
with
velocity  $\vec{v}$, this becomes
\be
Q(\vec{v};\vec k) =
\sqrt{\frac{1+v_0}{2}}(1+\frac{\vec{\alpha}.\vec{v}}{1+v_0})
Q(\vec{v}=0;\vec k)
\ee
\be
=\sqrt{\frac{1+v_0}{2}}
\pmatrix{1+ \frac{\vec{\sigma}\cdot\vec{v} \vec{\sigma}\cdot\vec
 k}{2m_Q(1+v_0)}\cr
\frac{\vec{\sigma}\cdot\vec{v}}{1+v_0}+\frac{\vec{\sigma}\cdot\vec
k}{2m_Q}\cr}
\ee
If the spectator is a single (anti)quark, the $S(\vec v;-\vec k)$ follows
from eq.16 by replacing $\vec k$ by $-\vec k$ and
$2m_Q$ by $2m_q$ (upper component) or $E_q+m_q$ (lower
component).
(For a
multiquark system the $S(\vec{v},-\vec{k})$ is a direct product of spinors
with appropriate momenta relative to the centre of momentum of the
system).
This way of writing the spinors for a boosted composite system is
essentially
 that
developed by Brodsky and Primack (see section 4 of ref
4) and establishes the identity with the HQET
decomposition at eq(11,12). Note the $\vec{\sigma}\cdot\vec{v}
\vec{\sigma}
\cdot\vec{k}$ term in the upper component of eqs (12, 16): this has
frequently
been overlooked in the literature and will be seminal in what follows.

The careful separation of the quark momentum within the hadron from
the
overall motion of the system is crucial when considering interactions that
 transfer momentum
(velocity) (such as spin dependent gluon exchange energy shifts in
spectroscopy
(ref 5) or the interactions with external currents, as in
electromagnetic  [4,6,7,8] or weak interactions). The use of the spinor eq(16)
and consistent
accounting of the spectator wavefunction $S(\vec v, -\vec k)$ in the
boosted
version of eq(13) most effectively leads to the correct interaction
Hamiltonian.

As an example consider a ``meson" consisting of a heavy quark, charge
$e_Q$, mass
$m_Q$, and a quasi-free  electrically neutral spectator fermion with mass
$m_q$.  The corresponding
electromagnetic interaction for this system is [4,6,8]

\bea
H_I & = & \left \{e_Q\frac {\vec p_Q\cdot\vec A_Q}{m_Q}
-\frac {e_Qg}{2m_Q}\vec \sigma_Q\cdot\vec B_Q-
\frac {e_Q}{2m_Q}(2g-1)
\vec \sigma_Q\cdot\left (\vec E_Q\times\frac
{\vec p_Q}{2m_Q}\right )\right \}\nonumber\\ &+&
\frac1{4M_T}\left(\frac{\vec
 \sigma_q}{m_q}-\frac{\vec
\sigma_Q}{m_Q}\right)\cdot(e_Q\vec E_Q
\times{\vec p_q})
\eea
where $\vec A_Q$, $\vec B_Q$ and $\vec E_Q$ are the
 electromagnetic fields at the position of the heavy quark.   We allow the
possibility
that $g\not = 1$ from the QCD effects.
The second line in Eq. 17 shows the crucial spectator-dependent or
``nonadditive terms" explicitly that are known to be essential in satisfying
the low energy theorems of Compton scattering[4,8].
They arise because

(i) When the spinor is written in the manner of eqs.(11,12) or (16)
it manifests correlations between the relative
(k) and overall(v) motion - specifically the
$\vec \sigma\cdot \vec v \vec \sigma\cdot \vec k$ upper component.
This
term in
 the spinor
of the ``active" quark (the one interacting with the external current)
generates
the $\vec \sigma_Q \vec E_Q$ term in the second line of eq17.

(ii)The momentum of the spectator(s) is conserved in a nontrivial
manner.
There
is a transfer from relative(k) to overall(v) momentum of the spectator(s)
induced by the recoil of the
system; this induces a contribution from the spectators to the Wigner
rotation of the hadron's overall spin. For a single light quark spectator,
 as in Eq17,
this spin rotation is given by
 $\bar S(\vec v',-\vec k')S(\vec v,-\vec k)$ (where
$\vec k - \vec k' = (M-m_Q)(\vec v - \vec v')$). This generates
the $\vec \sigma_q \vec E_Q$ term in the second
 line of eq17 and calls into question
the traditional assumption that the spectators are ``spin-
inert".\footnote{A
 covariant representation of states (e.g. refs 16,17) incorporates
 the spectator
spin-rotation correctly at leading order but it is not immediately
transparent
how the required spin-rotation arises in that formalism. The utility of our
detailed approach is that it shows how the change in overall velocity arises
as
a
 trade-off with relative momentum and enables contact to be made
between
the covariant approach, generalised to O(1/M), and explicit models.}

The result is that the vector current $J_{\mu}=(J_0,\vec J)$ for the heavy
quark
 $m_Q$ in
terms of the velocity $v$ is (to $0((v-v^\prime)^2\;;\;\vec{v}^3$)
\be
\vec J=\frac 12 (\vec v+\vec v^\prime) -\frac {i gM}{2 m_Q}\vec
\sigma_Q
 \times (\vec v^\prime -\vec v) -\frac i8(v^\prime_0-v_0)\left (
 \frac {2gM}{m_Q}\vec\sigma_Q  -\vec\sigma_T\right )
\times (\vec v^\prime+\vec v) +\vec
 J(\vec k)
\ee
and
\be
J_0=\frac 12 (v_0+v^\prime_0)-\frac i8\left (
 \frac {2gM}{m_Q}\vec\sigma_Q  -\vec\sigma_T\right )
\cdot\left [(\vec v^\prime+\vec v) \times (\vec
 v^\prime-\vec v)\right ]+J_0(k)
\ee
where
\be
\vec J(\vec k)=\frac {\vec k^\prime+\vec k}{2m_Q}-
\frac {i}4 (v^\prime_0-v_0)
\left \{\left [\frac {M}{m_Q}(2g -1)-1\right ]
\vec \sigma_Q\times \frac {\vec k^\prime+\vec k}{2m_Q}+
 \vec \sigma_q\times \frac {\vec k^\prime+\vec k}{2m_q}\right \},
\ee
\be
J_0(k)= \frac {k_0^\prime +k_0}{2m_Q}-\frac {i}4
\left \{\left [\frac {M}{m_Q}(2g -1)-1\right ]
\vec \sigma_Q\times  \frac {\vec k^\prime+\vec k}{2m_Q}
 +\vec \sigma_q\times \frac {\vec k^\prime+\vec k}{2m_q}\right
\}\cdot
(\vec v^\prime -\vec v)
\ee
are functions  of the relative momentum $\vec k$,
 and $\vec \sigma_T=\vec \sigma_Q+\vec \sigma_q$.
(For a flavour changing vector current between initial(i) and final(f)
states, the $\frac{M}{m_Q}$ in leading order becomes the average, namely,
$\frac{1}{2}(\frac{M_i}{m_Q}_i + \frac{M_f}{m_Q}_f)$; there are also
terms
proportional to $(m_i - m_f)$ at order $1/M$ which we shall not discuss
in
this paper).

 The nonadditive
 terms in Eq. 17 are crucial in generating the correct overall Wigner
rotation
proportional to $\vec \sigma_T$ in Eqs.18,19; they summarise the
kinematic
spin rotation associated with a boost from the quark to hadron overall
c.m. frame. The spectator
system does not simply act as a spin-neutral system
 (contrast e.g. ref. 9 and 16).
 Thus even though a flavor changing weak
 decay at first sight involves only the heavy
 active quark, we see that at non-zero
velocity transfer the {\bf light quark spectators do not trivially factor
out in the effective Lagrangian of the HQET and
 can play a dynamical role even
when $M_Q \to \infty$}.

The careful accounting of Wigner
 rotations is important when studying polarisation [10] at large
momentum
transfer. Examples of immediate relevance include the polarisation
[10] of vector mesons $D^*,K^*$ in semi-leptonic decays such as
$B\rightarrow D^* e\nu, D\rightarrow K^*e\nu$ and of the
$\Lambda_c, \Lambda_s$ in their baryonic analogues
[11]: indeed, the importance
of such care in maintaining covariance in
$B\rightarrow D^* e\nu$ has already been noticed in an explicit model
[12].
Note, in particular, that the $\vec{\sigma}_T$ in eq.18,19 does not
contribute
to the transition $B\rightarrow D^*$; if this had been incorrectly written
 as $\vec{\sigma}_Q$ then the term in parenthesis in eq18,19 would have
its
strength underestimated by (roughly) a factor of two.
We find that the relationships among form factors for
$\Lambda_b\rightarrow
\Lambda_c e\nu$ at $O(1/m)$ (ref 13,14) are unaffected in the
approximation
that $\omega =m_Q$, essentially because the light quark spectator system
has
no net spin.

The explicit appearance of light quark dynamics might appear to frustrate
some of the hopes that HQET makes clean cut statements about processes
involving heavy quark
hadrons.  However, the form of Eqs.18 to 21 and the physical
interpretation
of the light-quark contributions, suggest that HQET can remain effective
if judicious choice of frame is made. For flavor changing transitions
involving
heavy quarks, the Wigner rotation subtelties may be bypassed in leading
order
by choosing to work in the particular frame where $\vec v=-\vec v'$ (this
frame
[15] would correspond to the Breit frame in the case $m_i=m_j$). The
terms
linear in $\vec k/m_q$ integrate to zero for transitions involving S-state
hadrons, but can give non-trivial contributions elsewhere (e.g.ref18).
In studies of
hadron spectroscopy for a two-body system one need only work in the
$\vec
v=0$
frame for these concerns to become academic[8]; however for baryons,
where
the
$v_{qq}=0$ frame differs in general from the $\vec v_{Qqq}=0$ frame,
these
problems are rather central, in particular for the P-wave and other excited
states[5]. In transitions where the light quarks are active, such as
 $H_1 \rightarrow H_2 +(\pi or \rho)$, there can be analogous spin
rotations
of
 the
spectator heavy quark which need to be taken into account.

In this note we have ignored interactions between the
 heavy quark and other constituents; at $O(1/m_Q)$ these can generate
dynamical spin couplings between heavy and light quarks in addition to
the ``kinematic" ones discussed here.
The incorporation of scalar or vector binding potential
 follows
the procedure developed in refs 7,8 (for the case of vector currents where
$m^i_Q=m^f_Q$);
application of these ideas to flavour changing transitions,  involving
vector
 and
axial currents, will be described in
detail elsewhere.

We are indebted to J.Flynn, J.Korner, C.Sachrajda and A.Wambach for
comments.
This work was supported in part
by the United States Department of Energy under contract
 DE-AS05-76ER0-4936.


\begin{thebibliography}{15}
\bibitem{ } N. Isgur, M. Wise, Phys. Lett. {\bf B232}, 113 (1989); {\bf B237},
 527
(1990).
\bibitem{ } H. Georgi, Phys. Lett. {\bf B240}, 447 (1990);
 N. Isgur, M. Wise ``Heavy Quark Symmetry", CEBAF-92-17.
\bibitem{ } M. J. Dugan, M. Golden and B. Grinstein, to be published in
 Phys. Lett. {\bf B }
\bibitem{ } S. Brodsky and J. Primack, Ann. Phys. {\bf 52}, 315 (1969).
\bibitem{ } F.E. Close and R.H. Dalitz, p 411 in ``Low and Intermediate
Energy
Kaon-Nucleon Physics", D. Reidel pub. 1981.
\bibitem{ } F.E. Close and L. Copley, Nucl. Phys. {\bf B19}, 477 (1970).
\bibitem{ } F.E. Close and Z.P. Li, Phys. Rev. {\bf D42}, 2194 (1990); ibid
 2207.
\bibitem{ } H. Osborn, Phys. Rev. {\bf 176}, 1514 (1968); ibid 1523; F.E. Close
 and
H. Osborn, Phys. Rev. {\bf D2}, 2127 (1970).
\bibitem{ } P. Lepage et al, Cornell Univ report CLNS 92/1136 (1992);
 F. Hussain et al. Phys. Lett. {\bf B249}, 295(1990).
\bibitem{ } B. Grinstein et al, Phys. Rev. Letters {\bf 56}, 298 (1986);
J. Anjos et al, Phys. Rev. Letters {\bf 62}, 722 (1989);
F. Gilman and R. Singleton, Phys. Rev.  {\bf D41}, 142 (1990).
\bibitem{ }R.Singleton, Phys Rev. {\bf D43}, 2939(1991);
F.Hussein and J. Korner, Z.Phys. {\bf C41},657 (1991).
F.E.Close et al,``Heavy Quark Theory and b-polarisation at LEP",
RAL-92-016, J.Phys.G (in press).
\bibitem{ } E. Golowich et al. Phys. Lett. {\bf B213}, 521 (1988).
\bibitem{ } H. Georgi et al, Phys. Lett.  {\bf B252}, 456 (1990).
\bibitem{ } J.G. Korner and M. Kramer, DESY 91-123 (1991).
\bibitem{ } N.Isgur, Phys Rev {\bf D43},810 (1991).
\bibitem{ } F.Hussein, J.Korner and G.Thompson,Ann.Phys {\bf 206},334
(1991).
\bibitem{ } A.Falk, H.Georgi,B.Grinstein and M.B.Wise, Nucl.Phys {\bf
B343},
1 (1990).
\bibitem{ } S.Balk et al.,``Covariant Trace Formalism for Heavy Meson S-
wave
to P-wave transitions", Univ Mainz report (1992)
\end{thebibliography}
\end{document}